\documentclass[showpacs,preprintnumbers,amsmath,amssymb]{revtex4}
\usepackage[dvips]{graphicx}
\usepackage{dcolumn}
\begin{document}

\title
{
High field superconducting phase diagrams including Fulde-Ferrell-Larkin-Ovchinnikov vortex states
}

\author{Ryusuke Ikeda}

\affiliation{%
Department of Physics, Kyoto University, Kyoto 606-8502, Japan
}

\date{\today}

\begin{abstract}

Motivated by a striking observation of an Fulde-Ferell-Larkin-Ovchinnikov (FFLO) vortex state in the heavy fermion material CeCoIn$_5$ in fields {\it perpendicular} to the superconducting planes (${\bf H} \parallel c$), superconducting phase diagrams including an FFLO state of uniaxially anisotropic superconductors are systematically studied. In the ballistic limit with no quasiparticle's (QP's) relaxation, the high field superconducting state in ${\bf H} \parallel c$ and in the low temperature limit should be not the FFLO state modulating along ${\bf H}$, appeared in CeCoIn$_5$, but a different vortex state with a modulation {\it perpendicular} to the field. It is argued that an enhancement near $H_{c2}(0)$ of the QP relaxation rate, presumably originating from a {\it nonsuperconducting} quantum critical fluctuation, in this material is the origin of the absence of the latter modulated state and of the strange ${\bf H} \parallel c$ phase diagram in which the FFLO state is {\it apparently} different from that in ${\bf H} \perp c$. 
\end{abstract}


%\keyword{}

\maketitle

\section{Introduction}

The recent discovery \cite{Bianchi} of a high field superconducting (SC) state in a uniaxially anisotropic heavy fermion superconductor CeCoIn$_5$ in fields parallel to the SC layers (${\bf H} \perp c$) has led to renewed interests in the Fulde-Ferrell-Larkin-Ovchinnikov (FFLO) state \cite{Pauli}. The identification between an FFLO state and the detected high field phase, accompanied by a discontinuous $H_{c2}$-transition \cite{Izawa}, is based on an indication of the strong paramagnetic effect in this material \cite{Izawa,JPCM} and on a derived vortex phase diagram including the discontinuous $H_{c2}$-transition \cite{AI}. Based on the {\it conventional} picture on FFLO states in the vortex-free Pauli limit, \cite{Pauli} however, the presence of an FFLO state in CeCoIn$_5$ seems to be an unexpected event in several respects, although this material seems to have a quasi two-dimensional (Q2D) electronic structure. First, one needs to clarify why an FFLO state has appeared in the material with a {\it weak} uniaxial anisotropy \cite{AI2}, although it has not been clearly observed until recently in {\it strongly} anisotropic Q2D materials. In particular, the recent observation of an FFLO state in fields {\it perpendicular} to the SC layers (${\bf H} \parallel c$) \cite{Kumagai,Brazil} is the most striking in this sense, because an FFLO state is conventionally expected not to appear in this configuration dominated by the orbital pair-breaking. The feature that a flat FFLO transition curve, usually expected in the vortex-free Pauli limit, was seen not in ${\bf H} \perp c$ but in ${\bf H} \parallel c$ remains to be explained \cite{Kumagai}. Second, an observed pressure-induced extension of the FFLO temperature region \cite{Dresden} is apparently inconsistent with the fact that an FFLO state has appeared in not a material well described by the weak-coupling model but CeCoIn$_5$ with strong electron correlation. 

%Results of a ultrasound measurement \cite{ISSP} are understood based on a reduc%tion of the tilt modulus in the FFLO state with a one-dimensional modulation {\%it parallel} to the field \cite{RItilt}.
In this paper, high field phase diagrams including FFLO vortex states of a superconductor with a Q2D electronic structure are systematically examined to explain the striking observations in CeCoIn$_5$ mentioned above on the same footing. Phase diagrams are discussed first in the ballistic limit, where the quasiparticle's (QP's) mean free path $l_{\rm QP}$ is infinitely long, and next by assuming a finite $l_{\rm QP}$ and a slight change of the shape of Fermi surface (FS). The phase diagrams, in both fields parallel and perpendicular to the layers, of the types realized in CeCoIn$_5$ are obtained only when a finite $l_{\rm QP}$ is assumed in a system with a moderately large Maki parameter $\alpha_{{\rm M}}$. Inclusion of a finite $l_{\rm QP}$ is motivated by two observations: One is the theoretical fact that, in contrast to the conventional ansatz \cite{GG}, the ground state just below $H_{c2}(0)$ in the ballistic limit is the FFLO state modulating {\it not} along ${\bf H}$ but in the plane perpendicular to ${\bf H}$ \cite{AI,RItilt}. The other is an experimental result suggesting a strong and nonmonotonous field-dependence of $l_{\rm QP}$: Recent two transport measurements \cite{Kasahara,Ono} have shown that $l_{\rm QP}$ at much lower temperatures than $T_c(H=0)$ is anomalously long in low enough fields (i.e., deep in the SC state) {\it and} in much higher fields than $H_{c2}(0)$, implying that it is the shortest near $H_{c2}(0)$. Actually, the pressure induced extension of the FFLO region \cite{Dresden} is convincingly explained as a consequence of the pressure dependences of $T_c$ {\it and} $l_{\rm QP}$.  

This paper is organized as follows. In sec.II, a theoretical derivation of formulae needed in numerically deriving a phase diagram is explained. Numerical results of phase diagrams and thermodynamic quantities are shown and discussed in comparison with experimental ones in sec.III. In sec.IV, the obtained results are further discussed. 

\section{Theoretical Method}

Our analysis starts from the quasi 2D weak-coupling BCS Hamiltonian with the Zeeman energy $\mu_B H$ and a $d$-wave pairing interaction which consists of the following three terms 
\begin{eqnarray}
{\cal H}_0 &=& d \sum_{\sigma, j}  \int d^2 r_{\perp} 
        ( \, {\varphi}_{j}^{\sigma}({\bf r_\perp}) \, )^\dagger \Bigg[ \, 
        \frac{ ({\rm -i}{\nabla}_{\perp} + e{\bf A} )^2}{2m_e} - \sigma \mu_B H \, 
        \Bigg] {\varphi}_{j}^{\sigma}({\bf r_\perp}), 
\label{BCSkin}
\end{eqnarray} 

%%%%%
\begin{eqnarray}
{\cal H}_J &=& - \frac{J \, d}{2} \sum_{\sigma, j}  \int d^2 r_{\perp} \Bigg(
{\varphi}_{j}^{\sigma \dagger}({\bf r_\perp})
{\varphi}_{j+1}^{\sigma}({\bf r_\perp}) +  {\varphi}_{j+1}^{\sigma \dagger}({\bf r_\perp}) {\varphi}_{j}^{\sigma}({\bf r_\perp}) \Bigg),
\label{interlayer}
\end{eqnarray}
%%%%%
and 
\begin{eqnarray} 
{\cal H}_{\rm int} &=& - \frac{|g| \, d}{2} \sum_{\sigma, j} 
 \int \frac{d^2 k_{\perp}}{(2 \pi)^2} 
	B_{\sigma, j}^{\dagger}({\bf k_\perp})
          B_{\sigma, j}({\bf k_{\perp}}).  
\label{BCSint}
\end{eqnarray}
Here, 
$B_{\sigma, j}({\bf k_{\perp}}) = \sum_{\bf p_\perp} {\hat \Delta}_{\bf p}
a_j^{-\sigma}({\bf -p_-}) a_j^{\sigma}({\bf p_+})$, ${\bf p}_\pm = {\bf p_\perp \pm {\bf k}_\perp}/2$, $j$ is the index numbering the SC layers, ${\bf p}_\perp$ is the component of ${\bf p}$ parallel to the layers, $d$ is the interlayer spacing, and $m_e$ is the effective mass of a 
quasi-particle. Further, ${\hat \Delta}_{\bf p}$ is the normalized orbital part of the pairing-function which, in the case of $d_{x^2-y^2}$-pairing, is written as $\sqrt{2}({\hat p}_x^2 - {\hat p}_y^2)$ in terms of the unit vector ${\hat {\bf p}}$ parallel to the layers. Hereafter, the gauge field ${\bf A}$ will be assumed to consist only of a term expressing the external field $H$, i.e., we work in the type II limit with no spatial variation of flux density, because we are interested mainly in the field region near $H_{c2}$. 
Further, $\sigma \mu_B H = \mu_B H$ or $-\mu_B H$ is the Zeeman energy. In discussing our calculation results, the strength of the paramagnetic effect under a field in the $j$-direction is measured by the Maki parameter $\alpha_{M,j} = \sqrt{2} H^{({\rm orb})}_j(0)/H_P(0)$. Here, $H_P(0) = \pi T_c/(\sqrt{2} e^{\gamma_{\rm E}} \, \mu_B) \simeq 1.2 T_c/\mu_B$ is the Pauli limiting field at $T=0$ defined within the weak-coupling BCS model, where $\gamma_{\rm E}=0.577$ is an Euler constant, while $H^{({\rm orb})}_j(0)$ is the orbital limiting field at $T=0$ for fields parallel to the $j$-direction. 

For the moment, the case with infinite $l_{\rm PQ}$ (the ballistic limit) will be considered in the weak-coupling approximation, and effects of a finite $l_{\rm QP}$ and strong correlation will be incorporated later. Unless specifically noted, the FFLO state on which we focus has a modulation {\it parallel} to ${\bf H}$. Hereafter, the expressions necessary for examining the $H$-$T$ phase diagram in perpendicular fields ${\bf H} \parallel c$ will be derived by closely following the methods used in Ref.\cite{RItilt} for ${\bf H} \perp c$ case, and hence, just the essential part of the formulation and main results in ${\bf H} \parallel c$ will be presented below. To describe the FFLO state modulating along ${\bf H} \parallel c \parallel {\hat z}$, we can focus on the $n=0$ Landau level (LL) modes of the SC order parameter $\Delta({\bf r})$ in equilibrium \cite{GG}, because the $n=0$ LL modes are isotropic in nature in the $x$-$y$ plane and thus, cannot accommodate an FFLO modulation in the $x$-$y$ plane. Then the FFLO state (more precisely, the LO state) is described as 
\begin{equation}
\Delta({\bf r}) = \sqrt{2} \, T_c \, \alpha_e \varphi_0(x,y) \, 
{\rm cos}(Qz), 
\end{equation}
where $\varphi_0(x,y)$ is the Abrikosov lattice solution formed in the $n=0$ LL under ${\bf H} \parallel {\hat z}$, and ${\bf Q} = Q {\hat z}$ is the wavevector of the FFLO modulation. Then, the mean field GL free energy density ${\cal F}$ in $n=0$ LL takes the form 
\begin{eqnarray}
\frac{\cal F}{N(0) T_c^2} &=& a_{0}(Q) \alpha_e^2 + \frac{V_4(Q)}{2} \alpha_e^4 + \frac{V_6}{3} \alpha_e^6 \nonumber \\
&=& c^{(0)}(\alpha_e) + c^{(2)}(\alpha_e) \, {\overline q}^2 + c^{(4)}(\alpha_e) \, {\overline q}^4 
\label{FE}
\end{eqnarray}
represented by the amplitude $\alpha_e$ and the FFLO order parameter ${\overline q}= 2 \pi Q \xi_0$ \cite{AI}. Here, $N(0)$ is the density of states per spin, $\xi_0=v_F/(2 \pi T_c)$ the in-plane coherence length, and $v_F$ the Fermi velocity in 2D case. Microscopic details are largely reflected in the expressions of these GL coefficients, $a_0$ and $V_m$. If necessary, the coefficients $a_0$ and $V_m$ may be expanded in powers of $Q^2$: 
\begin{eqnarray}
a_0(Q) &=& a_0(0) + a_0^{(2)} {\overline q}^2 - a_0^{(4)} {\overline q}^4, \nonumber \\
V_4(Q) &=& V_4(0) - V_4^{(2)} {\overline q}^2 + V_4^{(4)} {\overline q}^4,
\label{GLcoef1}
\end{eqnarray}
where the index "$0$" of $a_{0}$ indicates the LL index. 
The ${\overline q}$-dependence of the GL coefficients will be kept up to the quartic term so that $c^{(2)}$ is given by $c^{(2)}= \alpha_e^2 ( a_0^{(2)} - \alpha_e^2 V_4^{(2)}/2)$. As stressed elsewhere \cite{RItilt}, inclusion of ${\overline q}$-dependence of $V_4$ is necessary to keep a stable FFLO state in ${\bf H} \perp c$. The same treatment will also be used in ${\bf H} \parallel c$. 

The onset temperature $T_0$ at which the mean field $H_{c2}$-transition becomes discontinuous is given as the position at which $V_4(Q_m)$ becomes negative upon cooling while $V_6 > 0$, where $Q_m$ is the equilibrium value of the wave number of the FFLO modulation, and a second order transition line $H_{\rm FFLO}(T)$ is determined as the line on which $c^{(2)}(\alpha_{\rm e})$ becomes negative on cooling, while $c^{(4)}(\alpha_{\rm e}) > 0$. The discontinuous $H_{c2}$-transition curve is determined by 
\begin{equation}
a_0(Q_m)=\frac{3}{16} \, \frac{(V_4(Q_m))^2}{V_6}.
\label{fot}
\end{equation}
Further, by minimizing ${\cal F}$ with respect both to $Q$ and $\alpha_e$, 
$\alpha_e^2$ is determined by  
\begin{eqnarray}
\alpha_e^2 (Q_m) &=& \frac{-V_4(Q_m) + \sqrt{(V_4(Q_m))^2 - 4 a_0(Q_m) V_6}}{2 V_6}, 
\label{equilalpha}
\end{eqnarray}
while  
\begin{eqnarray}
{\overline q}^2_m \equiv (2 \pi Q_m \xi_0)^2 = \frac{-a_0^{(2)} + V_4^{(2)} (\alpha_e(Q_m))^2/2}{2(-a_0^{(4)}+(\alpha_e(Q_m))^2 V_4^{(4)}/2)},
\label{equilq} 
\end{eqnarray}
if $a_0^{(2)} - V_4^{(2)} (\alpha_e(Q_m))^2/2 < 0$, and ${\overline q}_m=0$ otherwise. 
For instance, in the case with a small but nonvanishing $q_m^2$, one obtains 
\begin{equation}
\alpha_e^2 \simeq \alpha_e^2(0) + \delta \alpha_e^2 \, {\overline q}^2_m
\end{equation}
up to O(${\overline q}_m^2$), 
where 
\begin{eqnarray}
{\overline q}^2_m \simeq \frac{-a_0^{(2)} + V_4^{(2)} \alpha_e^2(0)/2}{2(-a_0^{(4)}+ \alpha_e^2(0) V_4^{(4)}/2 - \delta \alpha_e^2 V_4^{(2)}/4)},
\end{eqnarray}
and 
\begin{equation}
\delta \alpha_e^2 = - \frac{1}{2 V_6} \biggl[ V_4^{(2)} + \frac{-V_4^{(2)} V_4(0) + 2 V_6 a_0^{(2)}}{\sqrt{(V_4(0))^2 - 4 a_0(0) V_6}} \biggr]. 
\end{equation}
In numerical calculations we have performed, we always find $\delta \alpha_e^2 < 0$. That is, the space average of $|\Delta|^2$ is reduced in entering the FFLO state by increasing $H$.

To derive eq.(\ref{FE}), the familiar route \cite{Popov,AI,RItilt} for deriving a GL action microscopically will be taken. Formally, the quadratic term of the GL expression is written as 
%%%%%
\begin{equation}
{\cal F}_2 
= 	\frac{1}{V} \int d^3 r {\Delta}^*({\bf r})
	\left( \frac{1}{|g|}-\hat{K}_{2}({\bf \Pi}) \right) 
	{\Delta}({\bf r}), 
\label{F21}
\end{equation}
%%%%%%
where 
%%%%%
\begin{eqnarray}
 \hat{K}_2 ({\bf \Pi}) &=& \frac{T}{2} \sum_{\varepsilon, \sigma} \int_{\bf p} |{\hat \Delta}_{\bf p}|^2 G_{\varepsilon, \sigma}({\bf p}) G_{-\varepsilon, -\sigma}(-{\bf p}+{\bf \Pi}),
\end{eqnarray}
${\bf \Pi}=-{\rm i} \partial/\partial{\bf r} + 2 e {\bf A}({\bf r})$, and 
\begin{equation}
G_{\varepsilon, \sigma}({\bf p})
=\Big[{\rm i}{\varepsilon} + \sigma \mu_B H - \varepsilon_{\bf p} \Big]^{-1} ,
\label{Green2}
\end{equation}
is the quasiparticle Green's function in the normal state in $H=0$, where $\varepsilon_{\bf p}$ is the single particle dispersion measured from the Fermi level. Just as in the semiclassical approach \cite{Eilenberger}, the details of FS will be assumed to be reflected just in the Fermi velocity vector ${\bf w}$ and the integral {\it on} the FSs when performing the momentum integral. Then, we have 
\begin{eqnarray}
\hat{K}_2 &=&
 \pi N(0) T \sum_{\varepsilon, \sigma} \biggl\langle |{\hat \Delta}_{\bf p}|^2 \frac{{\rm i} {\rm sgn}(\varepsilon_n)}{2({\rm i}\varepsilon_n + \sigma \mu_B H) - {\bf w}\cdot{\bf \Pi}} \biggr\rangle_{\rm FS}  \nonumber \\ 
 &=& N(0) \int_0^\infty d\rho \, f(\rho) \biggl\langle |{\hat \Delta}_{\bf p}|^2 \exp({\rm i} \, T_c^{-1} \rho {\bf w}\cdot{\bf \Pi}) \biggr\rangle_{\rm FS},
\nonumber \\ \label{K21}
\end{eqnarray}
where 
\begin{equation}
f(\rho) = \frac{2 \pi \, t}{{\rm sinh}(2 \pi t \rho)} \, {\rm cos}\biggl(\frac{2 \mu_B H \, \rho}{T_c} \biggr), 
\label{f}
\end{equation}
$t = T/T_c$, and $\langle \,\, \rangle_{\rm FS}$ denotes the average over FS. 

Similarly, the 4-th order (quartic) term and the 6-th order one of the GL free energy density are written as 
\begin{eqnarray}
{\cal F}_4 &=& \frac{1}{2 V} \int d^3r \, {\hat K}_4({\bf \Pi}_j) \, \Delta^*({\bf r}_1) \, \Delta^*({\bf r}_3) \, \Delta({\bf r}_2) \, \Delta({\bf r}_4)|_{{\bf r}_j \to {\bf r}}, \nonumber \\ 
{\cal F}_6 &=& \frac{1}{3 V} \int d^3r \, {\hat K}_6({\bf \Pi}_j) \, \Delta^*({\bf r}_1) \, \Delta^*({\bf r}_3) \, \Delta^*({\bf r}_5) \, \Delta({\bf r}_2) \, \, \Delta({\bf r}_4) \, \Delta({\bf r}_6)|_{{\bf r}_j \to {\bf r}}, 
\label{F4}
\end{eqnarray}
where ${\bf \Pi}_j=-{\rm i} \partial/\partial{\bf r}_j 
+ 2 e {\bf A}({\bf r}_j)$. For instance, ${\hat K}_4$ is given by
\begin{eqnarray}
{\hat K}_4 &=& \frac{T}{2} \sum_{\varepsilon,\sigma} \int_{\bf p} |{\hat \Delta}_{\bf p}|^4 \, G_{\varepsilon, \sigma}({\bf p}) \, G_{-\varepsilon, -\sigma}(-{\bf p}+{\bf \Pi}_1^*) G_{-\varepsilon, -\sigma}(-{\bf p}+{\bf \Pi}_2) G_{\varepsilon, \sigma}({\bf p}+{\bf \Pi}_3^* - {\bf \Pi}_2) \nonumber \\ 
&=& 2 \pi N(0) T \sum_{\varepsilon, \sigma} \biggl\langle \frac{-{\rm i} {\rm sgn}(\varepsilon) \, |{\hat \Delta}_{\bf p}|^4}{d_1 d_2 d_3} 
\biggr\rangle_{\rm FS} \nonumber \\
&=& \frac{2}{T_c^2} \, N(0) \int\Pi_{j=1}^3 d\rho_j \, f\bigl(\sum_{j=1}^3 \rho_j \bigr) \biggl\langle |{\hat \Delta}_{\bf p}|^4 \, \exp\biggl[\frac{\rm i}{T_c}( \, \rho_1 {\bf w}\cdot{\bf \Pi}_1^* + \rho_2 {\bf w}\cdot{\bf \Pi}_2 + \rho_3 {\bf w}\cdot{\bf \Pi}_3^* \, ) \biggr] \biggr\rangle_{\rm FS}.
\label{K41}
\end{eqnarray}
The corresponding expression of ${\hat K}_6$ is obtained in the same manner \cite{RItilt}. 

Applying the parameter integrals used in describing ${\hat K}_2$ to obtaining the quartic and 6-th order terms \cite{AI,RItilt} and performing the operation \cite{RItilt} $\exp({\rm i} \, \rho T_c^{-1}{\bf w}\cdot{\bf \Pi}) \Delta({\bf r})$, we obtain 
\begin{eqnarray}
a_n(0) &=& \frac{1}{2}{\rm ln}(h) + \int_0^\infty d\rho \biggl[ \frac{1}{\rho} \exp\biggl(-\frac{\pi^2 \xi_0^2 \rho^2}{r_H^2} \biggr) - f(\rho) \biggl\langle |{\hat \Delta}_p|^2 \, {\rm L}_n(|{\overline \mu}|^2 \rho^2) \, \exp\biggl(-\frac{|{\overline \mu}|^2 \rho^2}{2} \biggr) \biggr\rangle_{\rm FS} \biggr], 
\nonumber \\
a_0^{(2)} &=& \frac{1}{2!} \int_0^\infty d\rho \, f(\rho) \rho^2 \biggl\langle \, \frac{w_z^2}{v_F^2} \, |{\hat \Delta}_p|^2 \exp\biggl(-\frac{|{\overline \mu}|^2 \rho^2}{2} \biggr) \biggr\rangle_{\rm FS}, \nonumber \\
a_0^{(4)} &=& \frac{1}{4!} \int_0^\infty d\rho  \, f(\rho) \rho^4 \biggl\langle \, \frac{w_z^4}{v_F^4} \, |{\hat \Delta}_p|^2 \exp\biggl(-\frac{|{\overline \mu}|^2 \rho^2}{2} \biggr) \biggr\rangle_{\rm FS}, \nonumber \\
V_4(0) &=& 3 \int_0^\infty \Pi_{j=1}^3 d\rho_j \, f\biggl(\sum_{j=1}^3 \rho_j \biggr) \biggl\langle |{\hat \Delta}_p|^4 \exp\biggl(-\frac{1}{2} \biggl(-\frac{1}{2} R_{24} + R_{14} \biggr) \biggr) \, {\rm cos}(I_4) \biggr\rangle_{\rm FS}, \nonumber \\
V_4^{(2)} &=& \frac{3}{2!} \int_0^\infty \Pi_{j=1}^3 d\rho_j \, f\biggl(\sum_{j=1}^3 \rho_j \biggr) \biggl(\sum_{j=1}^3 \rho_j^2 - \frac{1}{3} \sum_{i \neq j} (-1)^{i+j} \, \rho_i \rho_j \biggr) \nonumber \\
&\times& \biggl\langle \, \frac{w_z^2}{v_F^2} \, |{\hat \Delta}_p|^4 \exp\biggl(-\frac{1}{2} \biggl(-\frac{1}{2} R_{24} + R_{14} \biggr) \biggr) \, {\rm cos}(I_4) \biggr\rangle_{\rm FS}, \nonumber \\
V_4^{(4)} &=& \frac{3}{4!} \int_0^\infty \Pi_{j=1}^3 d\rho_j \, f\biggl(\sum_{j=1}^3 \rho_j \biggr) \biggl[ \, \sum_{j=1}^3 \rho_j^4 + \sum_{i \neq j} ( 3 \rho_i^2 \rho_j^2 - 2 (-1)^{i+j} \rho_i \rho_j (\rho_{6-i-j})^2 - \frac{4}{3}(-1)^{i+j} \rho_i \rho_j^3) \biggr] \nonumber \\
&\times& \biggl\langle \, \frac{w_z^4}{v_F^4} \, |{\hat \Delta}_p|^4  \exp\biggl(-\frac{1}{2} \biggl(-\frac{1}{2}R_{24} + R_{14} \biggr) \biggr) \, {\rm cos}(I_4) \biggr\rangle_{\rm FS}, \nonumber \\
V_6 &=& -15  \int \Pi_{j=1}^5 d\rho_j \, f\biggl(\sum_{k=1}^5 \rho_k \biggr) \biggl\langle |{\hat \Delta_p}|^6 \exp\biggl( -\frac{1}{2}(R_{16}+R_{26}) \biggr) \, {\rm cos}(I_6) \biggr\rangle_{\rm FS},
\label{GLcoef2}
\end{eqnarray}
where $h=H/H^{({\rm orb})}_{\rm 2D}(t\!\!=\!\!0) = 3.57 \, e \, \xi_0^2 H$, $H^{({\rm orb})}_{\rm 2D}$ is the 2D limit of the orbital limiting field $H^{({\rm orb})}_c$ in ${\bf H} \parallel c$, ${\rm L}_n(x)$ is the $n$-th order Laguerre polynomial, 
\begin{eqnarray}
R_{14} &=& |{\overline \mu}|^2 
(\sum_{j=1}^3 \rho_j^2 + \rho_2(\rho_3+\rho_1)), 
\nonumber \\
R_{24} &=& {\rm Re}({\overline \mu}^2) (\rho_2^2 + (\rho_3-\rho_1)^2), \nonumber \\
I_4 &=& \frac{{\rm Im}({\overline \mu}^2)}{4} (\rho_2^2 - (\rho_3-\rho_1)^2), \nonumber \\
R_{16} &=& |{\overline \mu}|^2 \biggl(e_1+e_2+e_3+\frac{2}{3}e_4 e_5 \biggr), \nonumber \\
R_{26} &=& {\rm Re}{\overline \mu}^2 \biggl(e_1+e_2+e_3-\frac{e_4^2+e_5^2}{3}-\frac{2}{3} (e_6+e_7+e_8+e_9) \biggr), \nonumber \\ 
I_6 &=& \frac{{\rm Im}({\overline \mu}^2)}{4} (e_1+e_2-e_3+\frac{e_5^2-e_4^2}{3} + \frac{2}{3}(e_8+e_9-e_6-e_7)) \nonumber \\ 
e_1 &=& (\rho_3+\rho_5)^2 + (\rho_3+\rho_4)^2, \nonumber \\
e_2 &=& (\rho_1+\rho_4+\rho_5)^2, \nonumber \\
e_3 &=& \rho_3^2 + \rho_4^2 + (\rho_2-\rho_5)^2, \nonumber \\
e_4 &=& \rho_1+2(\rho_3+\rho_4+\rho_5), \nonumber \\
e_5 &=& \rho_2-\rho_3-\rho_4-\rho_5, \nonumber \\
e_6 &=& (\rho_4-\rho_5)^2 + (\rho_1+\rho_5-\rho_3)^2, \nonumber \\
e_7 &=& (\rho_1+\rho_4-\rho_3)^2, \nonumber \\
e_8 &=& (\rho_3-\rho_4)^2 + (\rho_2+\rho_3-\rho_5)^2, \nonumber \\
e_9 &=& (\rho_2+\rho_4-\rho_5)^2,
\end{eqnarray}
and 
\begin{equation}
{\overline \mu}= \frac{\sqrt{2} \pi \xi_0}{r_H v_F}(w_x + {\rm i} w_y). 
\end{equation}

%%%%%%%%%%%%%%%%%%%
\begin{figure}[t]
\scalebox{1.2}[1.2]{\includegraphics{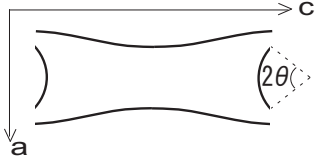}}
\caption{Cross section in $c$-$a$ plane of the model FS (solid curves) composed of a corrugated cylinder and a small portion with large $|w_c/w_{ab}|$ modelled as a piece of sphere. Any anisotropy in the $a$-$b$ plane is neglected here. } \label{fig.1}
\end{figure}
%%%%%%%%%%%%%%%%%

To concretely perform the average $\langle \,\,\,\, \rangle_{\rm FS}$ in the above expressions of the GL coefficients, an appropriate FS needs to be chosen. 
To explain the appearance \cite{Kumagai} of the FFLO state in ${\bf H} \parallel c$, mentioned in Introduction, in CeCoIn$_5$ with a Q2D structure, the use of a simple cylindrical FS is not appropriate: In the case of the purely cylindrical FS with a corrugation, the FFLO modulation does not become parallel to ${\bf H} \parallel c$ \cite{RIM2S} in spite of the field configuration dominated by the orbital depairing, reflecting that such a modulation tends to occur in a direction with the largest Fermi velocity \cite{MN}. This is not in conflict with the presence of the FFLO phase in CeCoIn$_5$ in ${\bf H} \parallel c$ because the FS sheet with the heaviest mass of quasipaticles in this material is not a pure cylinder with a corrugation but accompanied by a small portion with a large $|w_{c}/w_{ab}|$. See the electron 14-th sheet of FS in Ref.\cite{Onuki} which has the heaviest effective mass and thus, is more effective for a $d$-wave superconductivity. In this work, a toy model of FS, sketched in Fig.1, is used in which the noncylindrical portion is incorporated as a small piece of the spherical FS with radius $k_F$, the Fermi wavenumber in 2D limit. 
The spanning angle $\theta$ (see Fig.1) measures the size of the noncylindrical portion inducing an FFLO modulation parallel to ${\bf H} \parallel c$, while the uniaxial anisotropy $\gamma_{an}$ of the coherence lengths arises mainly from the corrugation of the cylindrical FS.

%%%%%%%%%%%%%%%%%%%
\begin{figure}[t]
\scalebox{0.3}[0.3]{\includegraphics{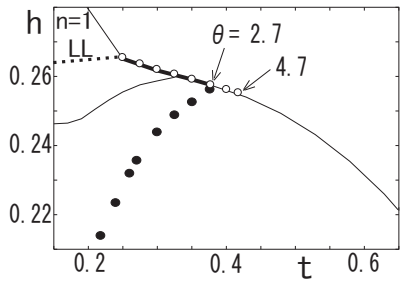}}
\caption{A typical $h$ v.s. $t$ phase diagram in ${\bf H} \parallel c$ composed of the mean field transition lines (solid curves) for $l_{\rm QP}=\infty$, $\alpha_{{\rm M}, c}=6.9$, and $\gamma_{an}=2.8$. Thin (thick) solid curves imply second order (discontinuous) transition lines for $\theta=2.7$ degrees. The $H_{c2}$-curve consists of three solid curves rising upon cooling, while the thin solid curve decreasing upon cooling is $H_{\rm FFLO}(T)$. The open (closed) symbols express the discontinuos $H_{c2}(T)$-transition line ($H_{\rm FFLO}(T)$-line) for $\theta=4.7$ degrees. The $H_{\rm FFLO}$-value for $\theta=4.7$ degrees does not decrease unlimitedly but saturates near 0.19 $H^{({\rm orb})}_{2D}(0)$ in $T \to 0$ limit. The $n=1$ LL vortex state occurs above the dotted curve. The onset $T_0$ of discontinuous $H_{c2}$-transition is indicated by an arrow for each $\theta$. Only the $H_{\rm FFLO}(T)$ curve was sensitive to such a small change of $\theta$-values. } \label{fig.2}
\end{figure}
%%%%%%%%%%%%%%%%%
%

\section{Possible Phase Diagrams and Thermodynamic Quantities}

A typical ${\bf H} \parallel c$ phase diagram obtained numerically in terms of the tools explained in sec.II is given in Fig.2, where $\gamma_{an}$ is the ratio $H^{({\rm orb})}_{ab}(0)/H^{({\rm orb})}_c(0)$, i.e., the anisotropy of the orbital limited $H_{c2}(0)$. The figure shows that a drastic shrinkage of the FFLO region occurs with decreasing $\theta$ (see Fig.1) reflecting the absence of the ${\bf H} \parallel c$ FFLO state in $\theta=0$ case \cite{RIM2S}. The high field ground state just below $H_{c2}(0)$ in the ballistic limit is formed here not in $n=0$ LL but in a higher LL, which is $n=1$ LL for $\alpha_{{\rm M}}$-values of our interest, and has some anisotropic inhomogenuity besides the vortices \cite{Klein}. The $n=1$ LL state has a striped structure \cite{Klein} due to nodal planes $\parallel {\bf H}$ and can be regarded as {\it another} FFLO state. In a GL free energy similar to eq.(1) but within the $n=1$ LL, its quartic term has a positive coefficient near $H_{c2}$ for the $\alpha_{{\rm M}}$-values of our interest here, and thus, a second order $H_{c2}$-transition occurs in $t < 0.25$ on the thin solid line following from $a_{1}(0)=0$ (see the first expression of eq.(\ref{GLcoef2})) and rising {\it steeply} on cooling. However, the presence of the $n=1$ LL state is inconsistent with the observations on CeCoIn$_5$ at low temperatures, in which no $H_{c2}$ curve rising steeply is seen, and the $H_{c2}$-transition remains discontinuous even at $t \simeq 0.015$ \cite{Brazil}. Note also that the change of FS flattens $H_{\rm FFLO}(T)$ line, while it does not affect the range of the $n=1$ LL state and the position of the $H_{c2}$-line. This suggests that other factors than the shapes of FS have to be taken into account to understand the ${\bf H} \parallel c$ phase diagram of CeCoIn$_5$. 

It should be stressed that this conclusion does not follow as far as trying to explain only the appearance of the FFLO state in ${\bf H} \perp c$ : As seen in Ref.\cite{IA04} where the elliptic FS was assumed, the FFLO state in $n=0$ LL manages to surmount an instability at low temperatures to the in-plane modulated state in $n=1$ LL for an appropriate FS because the discontinuous $H_{c2}(T)$ line in ${\bf H} \perp c$ shows a more remarkable rise on cooling than that in ${\bf H} \parallel c$. In contrast, it is difficult in ${\bf H} \parallel c$ to, in the ballistic limit, protect the $n=0$ LL state from the $n=1$ LL one.  

%%%%%%%%%%%%%%%%%%%
\begin{figure}[b]
\scalebox{0.3}[0.3]{\includegraphics{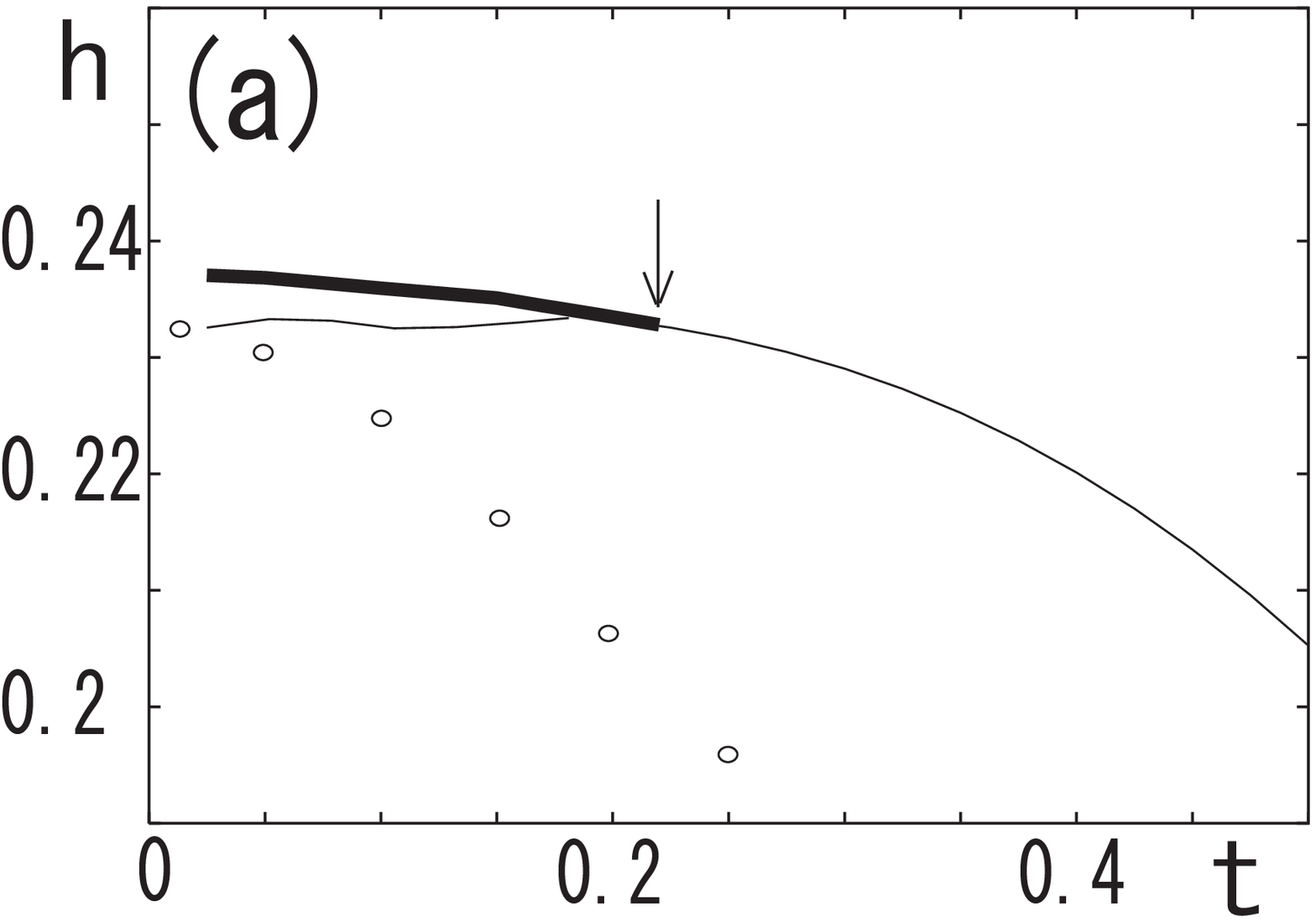}}
\scalebox{0.3}[0.3]{\includegraphics{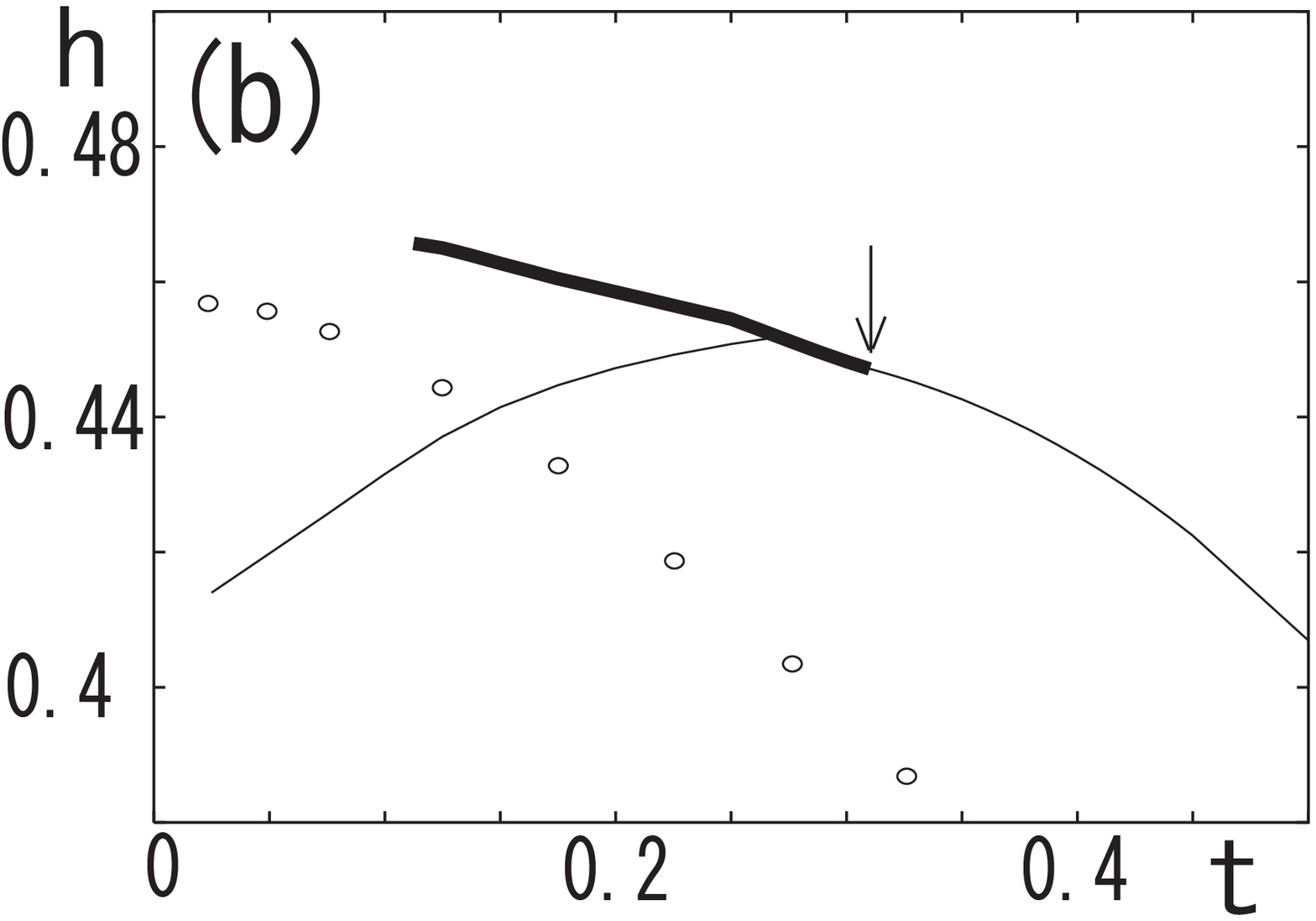}}
\caption{(a) Phase diagram in perpendicular fields (${\bf H} \parallel c$) in the case with a finite mean free path $l_{\rm QP}=15.5 \xi_0$. The values $\gamma_{an}=2.8$, $\theta=4.7$ degrees, and $\alpha_{{\rm M}, c}=6.9$ are used. The open circles indicate the curve defined by $a_1(0)=0$. The arrow indicates $T_0$. (b) The corresponding one under a parallel field in an antinodal direction following from $\alpha_{{\rm M}, \, ab}=7.69$ and the same values of $l_{\rm QP}$, $\theta$, and $\gamma_{an}$ as in (a). } \label{fig.3}
\end{figure}
%%%%%%%%%%%%%%%%%

%%%%%%%%%%%%%%%%%%%
\begin{figure}[t]
\begin{tabular}{cc}
\begin{minipage}{0.5\hsize}
\begin{center}
\scalebox{0.3}[0.3]{\includegraphics{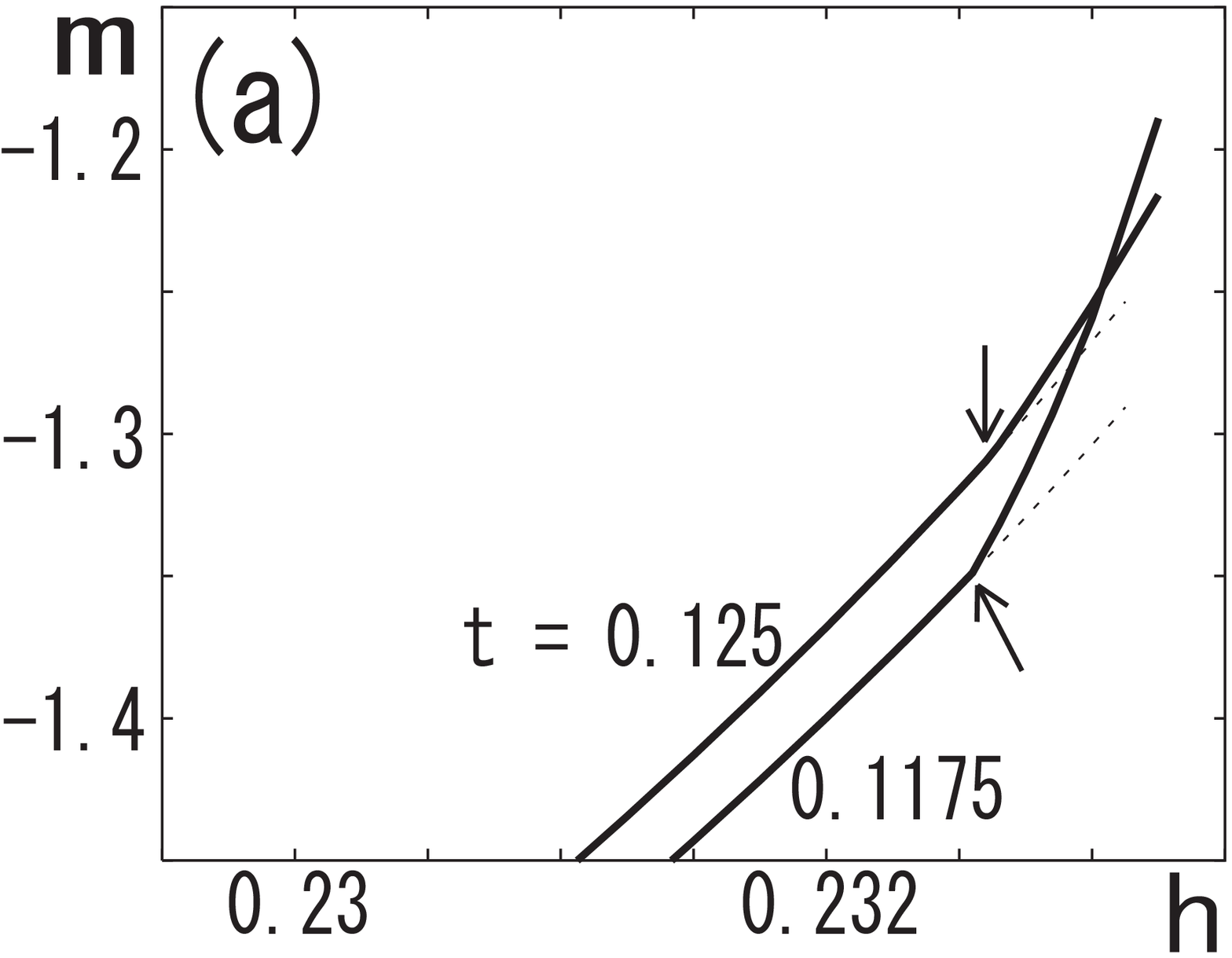}}
\end{center}
\end{minipage}
\begin{minipage}{0.5\hsize}
\begin{center}
\scalebox{0.3}[0.3]{\includegraphics{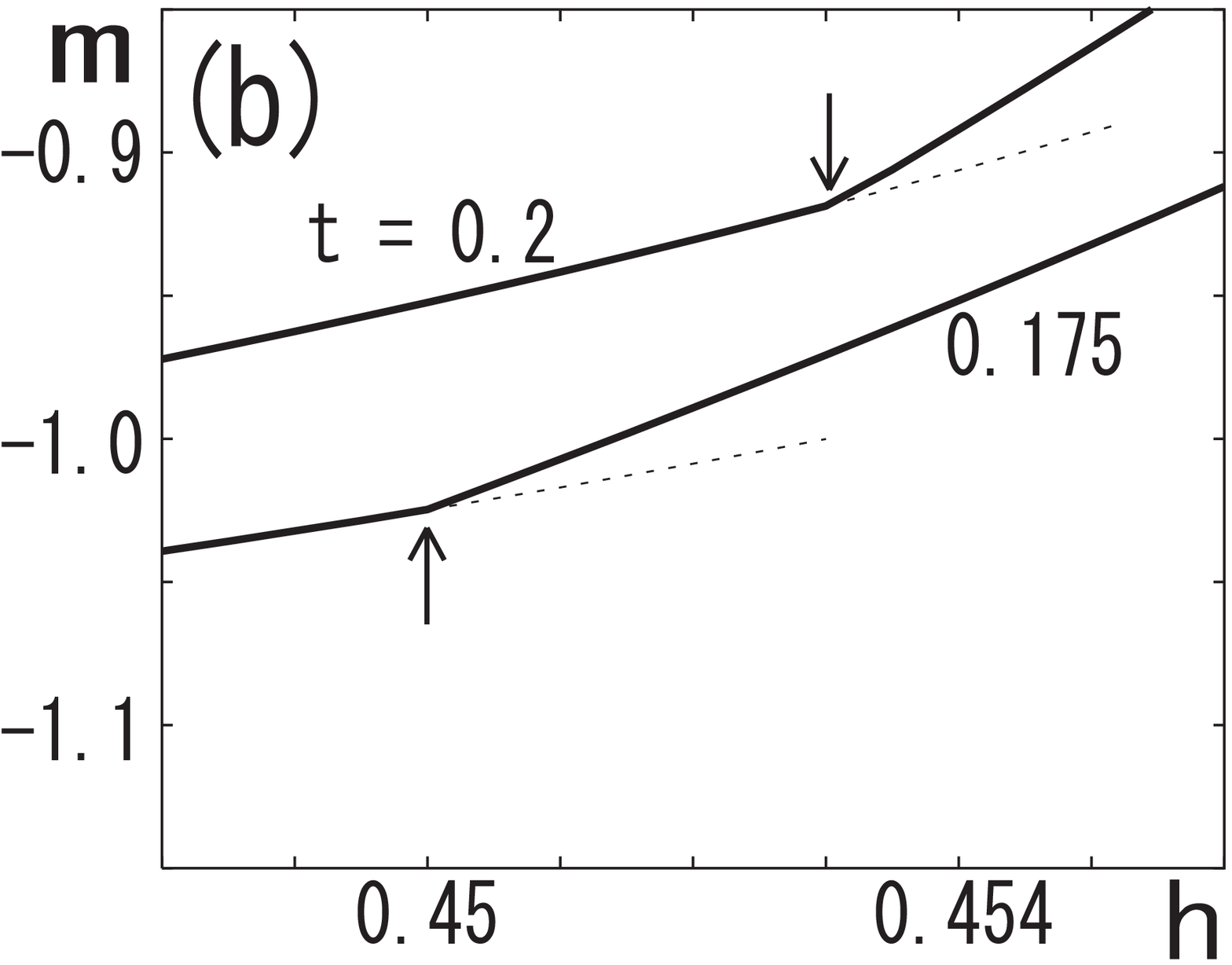}}
\end{center}
\end{minipage}
\end{tabular}
\caption{Field dependences of a dimensionless magnetization $m$ (see the text) in ${\bf H} \parallel c$ (a) and ${\bf H} \perp c$ (b) corresponding to Fig.3(a) and (b), respectively. Each arrow indicates the corresponding $H_{\rm FFLO}$. 
} \label{fig.4}
\end{figure}
%%%%%%%%%%%%%%%%%%%%% 

Next, effects of a finite QP's lifetime or $l_{\rm QP}$ will be considered by neglecting possible $h$ dependences of $l_{\rm QP}$ for the moment. A crucial role of a $H$-dependence of $l_{QP}$ will be pointed out in sec.IV. Possible origins of such a finite $l_{\rm QP}$ in CeCoIn$_5$ are an impurity scattering and some magnetic fluctuation accompanying, if any, a quantum critical point. Although we focus here on the case of an impurity, the present analysis is qualitatively applicable to the case with a magnetic fluctuation as far as the finite QP damping is their main effect on QPs. The primary consequence of an impurity scattering is a finite QP's relaxation rate $v_F/l_{\rm QP}$ in anisotropic superconductors satisfying $\langle {\hat \Delta}_p \rangle_{\rm FS}=0$, where impurity-induced vertex corrections to $a_n$ and $V_m$ are negligible \cite{AI}. The QP relaxation rate is incorporated in eq.(\ref{Green2}) merely with replacement, $2|\varepsilon| \to 2|\varepsilon| + v_F/l_{\rm QP}$, where $\varepsilon$ is a Fermion Matsubara frequency. In the expressions of eq.(\ref{GLcoef2}), this replacement is represented \cite{AI} simply by replacing $f(\sum \rho)$ there with 
\begin{equation}
{\overline f}\biggl(\sum \rho \biggr) = \exp\biggl(-2 \pi \frac{\xi_0}{l_{\rm QP}} \, \sum \rho \biggr) \, f\biggl(\sum \rho \biggr).
\label{fbar}
\end{equation}
A typical example of phase diagrams following from the resulting GL free energy is given in Fig.3. 
In ${\bf H} \parallel c$, effects of the finite $l_{\rm QP}$ are stronger, and the FFLO state easily shrinks, while the corresponding state in ${\bf H} \perp c$, as in Fig.3 (b), survives over a broad field range keeping a downward transition curve (i.e., $d H_{\rm FFLO}/dT > 0$). It is a reflection of the fact that the ${\bf H} \parallel c$ FFLO state, supported by a small piece of FS, is fragile and may be easily destroyed by a weak perturbation. Such a stronger effect of the finite $l_{\rm QP}$ on the ${\bf H} \parallel c$ FFLO state is not surprising once recalling that the impurity-induced pinning of vortices in Q2D vortex states is much weaker in the parallel fields. In general, as the FFLO state shrinks via a change of FS or a finite $l_{\rm QP}$, the $H_{\rm FFLO}(T)$ curve tends to change from a downward curve with positive $d H_{\rm FFLO}/dT$ to a flat or upward one. 
More importantly, {\it another} modulated state in $n=1$ LL accompanied by a steep $H_{c2}$-curve was pushed down to $T=0$ and was lost by including a finite $l_{\rm QP}$ in both field configurations of Fig.3. It suggests that a finite QP damping is needed to obtain phase diagrams consistent with those of CeCoIn$_5$. We note that the $l_{\rm QP}$-value used in Fig.3 is comparable with the estimated one from thermal conductivity data in $H \geq 0.8$ (T) \cite{Kasahara}. 

We note that the above-mentioned disappearance, induced by a finite $l_{\rm QP}$, of the $n=1$ LL modulated state is a consequence of an increase of the LL splitting and thus, cannot be found based on the approach \cite{AK} neglecting the vortices, i.e., in the {\it limit} of vanishing LL splittings. In fact, in contrast to our results \cite{AI} and the observation in CeCoIn$_5$ \cite{Bianchi,Kumagai}, no phase diagrams including a discontinuous $H_{c2}$-transition are obtained for $d$-wave pairing systems with a finite $l_{\rm QP}$ in such a limiting case \cite{AK}. 

To understand features of the Abrikosov to FFLO transition at $H_{\rm FFLO}$, the magnetization $M$ and the specific heat in ${\bf H} \perp c$ have been calculated around $H_{\rm FFLO}$. The specific heat jump at $H_{\rm FFLO}$ (or $T=T_{\rm FFLO}$) is given by 
\begin{equation}
\frac{\Delta C_{\rm FFLO}}{T_{\rm FFLO}(H)} = \frac{N(0)}{2 c^{(4)}} \, \biggl(\frac{\partial c^{(2)}}{\partial t} \biggr)^2 \simeq \frac{\Delta C(0)}{20 \, c^{(4)} \, T_c} \biggl(\frac{\partial c^{(2)}}{\partial t} \biggr)^2, 
\end{equation}
in terms of the jump value $\Delta C(0)$ at $T_c$ in zero field. Calculations leading to Fig.3(b) show that $\Delta C_{\rm FFLO}/T_{\rm FFLO}$ is $0.051 \, \Delta C(0)/T_c$ for $t=0.175$ and $0.034 \, \Delta C(0)/T_c$ for $t=0.075$, respectively, which are, up to the factor $2$, in agreement with the values, $(0.065 \sim 0.09) \Delta C(0)/T_c$, estimated from the data \cite{Bianchi,JPCM}. The decrease of $\Delta C_{\rm FFLO}/T_{\rm FFLO}$ upon cooling (see also Fig.3 in Ref.\cite{Bianchi}) implies that, as Fig.4 (b) also shows, this transition becomes more continuous as the 
paramagnetic depairing is more effective upon cooling. 

In Fig.4, 
the FFLO transitions in the two field configurations are compared with each other through results of the normalized magnetization $m(T,H) \equiv 8 \pi \kappa^2 \, M /(0.12 H^{({\rm orb})}_{\rm 2D}(0))$, where $\kappa$ is the GL parameter (i.e, the ratio between the penetration depth and the coherence length) defined in low fields. Noting that $\partial {\cal F}/\partial \alpha_e = \partial {\cal F}/\partial Q = 0$ in equilibrium, the magnetization $M$ is given simply by 
\begin{eqnarray}
M &=& - \, N(0) T_c^2 \alpha_e^2 \biggl[ (a_0(0))' + \frac{\alpha_e^2}{2}(V_4(0))' + \frac{\alpha_e^4}{3}V'_6 + {\overline q}_m^2 \biggl( (a_0^{(2)})' \nonumber \\
&-& \frac{\alpha_e^2}{2} (V_4^{(2)})' \biggr) - {\overline q}_m^4 \biggl( (a_0^{(4)})' - \frac{\alpha_e^2}{2} (V_4^{(4)})' \biggr) \biggr],
\end{eqnarray}
where the prime implies the derivative with respect to $H$ (i.e., $(a_0(0))' = \partial a_0(0)/\partial H$). As mentioned in sec.II, the spatial average of $|\Delta|^2$ decreases due to the appearance of the FFLO modulation. In fact, this decrease of $|\Delta|^2$ is the origin of the slope changes at $H_{\rm FFLO}$ in thermal conductivity \cite{Capan} and penetration depth \cite{Martin} data. Consistently with this $|\Delta|^2$-decrease, $|m|$ also shows an additional reduction, appeared as a kink, on entering the FFLO state from below. Note that the kink at $H_{\rm FFLO}$ in ${\bf H} \parallel c$ becomes more remarkable rather at lower $t$ in contrast to the tendency in ${\bf H} \perp c$ mentioned above. It implies that a transition to a more fragile FFLO state is sharper, reflecting a rapid growth of ${\overline q}$ near $H_{\rm FFLO}$, and may become discontinuous \cite{kapncs} for smaller values of $l_{\rm QP}$ and/or $\theta$. Thus, for a more fragile FFLO state, $|m|$ decreases more rapidly with increasing $H$ through $H_{\rm FFLO}$. This feature is consistent with a qualitative difference seen between available $m$-data in ${\bf H} \parallel c$ and ${\bf H} \perp c$ \cite{Brazil,Tayama}. Further, a sharper change at $H_{\rm FFLO}$ of $m$ tends to reduce the magnetization jump at $H_{c2}$. Since a weaker discontinuous $H_{c2}$-transition should occur closer to the {\it virtual} second order $H_{c2}$-transition line (i.e., the extrapolations to lower temperatures of the upper thin solid curves in Fig.3), the real $H_{c2}$-line in such a case will become flatter at lower temperatures. This is a qualitative explanation on differences between the $H_{c2}(T)$ curves in ${\bf H} \perp c$ and ${\bf H} \parallel c$. Although it is a conventional wisdom that, when the $H_{c2}$-transition is of second order, an FFLO modulation increases the $H_{c2}$ value, the coexistence of an FFLO state with a flat $H_{c2}$ curve in CeCoIn$_5$ seems to be a consequence of the {\it discontinuous} $H_{c2}$ transition. 

\section{Discussions}

Finally, effects of the electron correlation will be considered here in relation to the $p$ (pressure) dependences of the phase diagrams \cite{Dresden}, in which $T_0$ and $T_{\rm FFLO}$ increasing with increasing $p$ are suggested. The mass enhancement of normal QPs, which is a main effect of electron correlation, is incorporated by replacing the Matsubara frequency $\varepsilon$ in eq.(\ref{Green2}) by $Z \varepsilon$, where $Z > 1$. Then, by neglecting an $\varepsilon$-dependence of $Z$, it is easily verified that the theoretical results in the preceding sections and the figures in sec.III, expressed via the {\it normalized} field $h$ and temperature $t$, remain unchanged under replacement 
\begin{eqnarray}
T_c &\equiv& T_c(1) \to T_c(Z), \nonumber \\
N(0) &\to& Z N(0), \nonumber \\
\xi_0 &\to& \xi(Z) = \frac{\xi_0 T_c}{Z T_c(Z)}, \nonumber \\
\frac{\Delta({\bf r})}{T_c} &\to& \frac{\Delta({\bf r})}{Z T_c(Z)}, 
\end{eqnarray}
and 
\begin{eqnarray}
\alpha_{{\rm M}} \to \alpha_{{\rm M}}(Z) \equiv \alpha_{{\rm M}} \, \xi_0 \frac{\mu_B(Z)}{\xi(Z) \mu_B}, 
\end{eqnarray}
where $\mu_B(Z) H$ is the Zeeman energy in the case with a mass enhancement included, and the dimensionless Fermi velocity ${\bf w}/v_F$ is unchanged. The $Z$ dependence in the above $\Delta$-replacement is familiar in the conventional formulation of the strong coupling superconductivity \cite{Schrieffer} and implies the difference between the energy gap and the anomalous selfenergy of QPs in the SC phase, while the $Z$ dependences of the density of states and the coherence length are simply due to the mass enhancement. In general, an enhanced electron correlation should increase $\mu_B(Z)$, $(\xi(Z))^{-1}$ and thus, $\alpha_{{\rm M}}(Z)$. On the other hand, we have verified that an increase of $\mu_B(Z)$, as expected, reduces $H_{c2}(0)$, while a decrease of $\xi(Z)$ results in an increase of $H_{c2}(0)$. Then, since $Z$ should decrease with $p$, the $p$-induced decrease of $H_{c2}(0)$ in ${\bf H} \parallel c$ \cite{Dresden,ISSP2} is mainly a reflection of an increase of $\xi(Z)$, while the $p$-induced increase of $H_{c2}(0)$ in the parallel fields \cite{Dresden,ISSP2} is a consequence of $p$-induced decrease of $\mu_B(Z)$ in ${\bf H} \perp c$ outweighing the increase of $\xi(Z)$. Then, the $p$-induced increase of $T_c(Z)$ \cite{Dresden} may be one possible origin of the strange $p$-induced {\it increases} \cite{Dresden} of the two temperature scales induced by the paramagnetic depairing, $T_{\rm FFLO}(H)$ and the onset $T_0$ of the discontinuous $H_{c2}$-transition. 

However, the $p$ dependence of the QP damping seems to be a more direct origin of those of the two temperature scales if, as suggested in Ref.\cite{Ono}, the main origin of the QP damping is a scattering via magnetic fluctuations created by the strong correlation and surviving in $t \to 0$ limit. In fact, as shown in Fig.3, we have to assume the presence of a small but finite scattering rate $\sim l_{\rm QP}^{-1}$ to reach similar phase diagrams to those observed in CeCoIn$_5$. According to a recent estimation of $l_{\rm QP}$ based on transport data \cite{Kasahara,Ono}, $l_{\rm QP}$ obtained by sweeping $h$ in the low $t$ region relevant to the FFLO phenomena seems to be the shortest near $H_{c2}(0)$. Then, the results in Figs.2 and 3 suggest the picture that, reflecting a weaker magnetic fluctuation at higher $p$, the resulting weaker QP damping at higher $p$ would lead to increases of the above-mentioned two temperature scales. Further, the strong $H$-dependence of $l_{QP}(H)$ suggested in Ref.\cite{Kasahara,Ono} seems to resolve a qualitative disagreement on the $H_{\rm FFLO}(T)$-line between the experimental phase diagrams \cite{Bianchi,Dresden,Capan,Martin} and Fig.3(b) in which $l_{QP}$ was assumed to be $H$-independent : The $H_{\rm FFLO}(T)$ curve closer to the $H_{c2}(T)$ line in Fig.3(b) shows a negative curvature ($d^2H_{\rm FFLO}/dT^2 < 0$) in contrast to the experimental one. However, if $l_{QP}$ remarkably decreases with increasing $H$ below $H_{c2}(T)$, the part of $H_{\rm FFLO}(T)$ closer to $H_{c2}(T)$ should be depressed so that $H_{\rm FFLO}(T)$ may get a positive curvature. Although discussing a microscopic picture of the magnetic fluctuation is beyond the scope of this work, the above argument implies that the presence of a magnetic or nonsuperconducting quantum critical point near $H_{c2}$, suggested \cite{Ono,Capan2} through transport measurements in the high field normal state, is the main origin of remarkable differences in the phase diagram from that expected in the weak coupling and clean (or ballistic) limit. Then, we speculate that, at higher pressures, the higher LL vortex state with a modulation, {\it perpendicular} to ${\bf H}$, of FFLO type might occur at low enough 
temperatures in 
CeCoIn$_5$ particularly in ${\bf H} \parallel c$. 

In summary, SC phase diagrams including an FFLO vortex state have been systematically examined to explain the $H$-$T$ phase diagrams of CeCoIn$_5$. By examining notable differences in the phase diagrams and thermodynamic properties seen between the two configurations, ${\bf H} \perp c$ and ${\bf H} \parallel c$, crucial roles of quasiparticle damping in the FFLO state have been pointed out. The origin of differences between {\it conventional} theoretical phase diagrams including an FFLO state and the observed one of CeCoIn$_5$ seems to consist in an enhanced quasiparticle damping presumably related to a magnetic quantum critical behavior near $H_{c2}$. 

\begin{acknowledgements}
The author is grateful to Y. Matsuda and C.F. Miclea for discussions. Numerical computations were carried out at YIFP in Kyoto University. This work is financially  supported by a Grant-in-Aid from the Ministry of Education, Culture, Sports, Science, and Technology, Japan. 
\end{acknowledgements}

%\section{References} 

\end{document}